# Energy- and Performance-Driven NoC Communication Architecture Synthesis Using a Decomposition Approach*

Umit Y. Ogras and Radu Marculescu

*Department of Electrical and Computer Engineering*
*Carnegie Mellon University*
*Pittsburgh, PA 15213-3890, USA*
*e-mail: {uogras,radum}@ece.cmu.edu*

*Abstract*

*In this paper, we present a methodology for customized communication architecture synthesis that matches the communication requirements of the target application. This is an important problem, particularly for network-based implementations of complex applications. Our approach is based on using frequently encountered generic communication primitives as an alphabet capable of characterizing any given communication pattern. The proposed algorithm searches through the entire design space for a solution that minimizes the system total energy consumption, while satisfying the other design constraints. Compared to the standard mesh architecture, the customized architecture generated by the newly proposed approach shows about 36% throughput increase and 51% reduction in the energy required to encrypt 128 bits of data with a standard encryption algorithm.*

## 1. Introduction

The main bottleneck in designing today's Systems-on-Chip (SoCs) comes from global interconnects. Besides the increasing complexity due to the growing number of devices on the same chip, global interconnects continue to cause severe synchronization errors, unpredictable delays and high power consumption. As a result, it has been suggested to replace these custom wires with structured on-chip networks [1-3].

Typical SoCs that implement the Network-on-chip (NoC) approach consist of a number of heterogeneous devices such as CPU or DSP cores, embedded memory and application specific components, that communicate using packet switching. The design of NoCs trades-off several important architectural choices, such as topology and routing strategy selection, mapping the target application to the network nodes, etc. We can conceive these architectural choices as representing a 3-D design space. The first dimension of this space is the design of the *communication infrastructure*, e.g. the topology of the network and the width of the channel links; this is analogous to designing the roads in a big city. The next degree of freedom comes from the selection of *communication paradigm* which can be based on deterministic, adaptive or stochastic routing strategies. This second dimension is analogous to following the actual paths, while driving in a city traffic. The final dimension is application *mapping* to the network nodes, which consists of placing the message source/sink pairs to network nodes with the objective of satisfying some design constraints (e.g. energy, performance). Consequently, mapping has a big impact on the communication traffic pattern.

Communication infrastructure is the usual starting point in the design process of an NoC. Due to simplicity, a regular (i.e. grid-like) topology is usually chosen and then, the application mapping and routing strategy selection are carried out concurrently to optimize one or more design constraints, such as energy [4-6]. The selection of the interconnect topology has a dramatic impact on the overall performance, area, and power consumption. Hence, constraining the network architecture to consider only regular topologies, while exploring the remaining two dimensions of the design space, produces sub-optimal operating points. Furthermore, varying sizes and shapes of the cores and large deviations in the communication requirements cause waste of silicon area and over-designed networks or performance bottlenecks [6,7]. Since exhaustive design space exploration is prohibitive, a new design methodology that considers all three design dimensions and copes with the inherent complexity is clearly needed.

To this end, we propose a methodology for communication-based customized topology synthesis. While there exists many different communication patterns in a network, certain generic communication primitives, such as gossiping (all-to-all communication), broadcasting (one-to-all) and multicasting (one-to-many) are encountered most frequently [10,11]. Our approach is based on decomposing the communication requirements' of the target application as a combination of communication primitives. After the decomposition step, these basic communication primitives are replaced by their optimal implementations. Finally, the customized topology is obtained by gluing the optimal implementations together, while satisfying the imposed design constraints.

We illustrate our methodology using some random benchmarks with various characteristics and a real application, the *Advanced Encryption Standard (AES)*. For comparison purposes, the AES algorithm is implemented using a customized architecture generated by the proposed algorithm and a standard mesh architecture. The direct comparison of these two designs using an FPGA prototype shows 36% throughput

*This research is supported in part by CyLab ARO under grant no. 9097 and by MARCO/DARPA Gigascale Systems Silicon Center under grant no. 2003-DT-660.

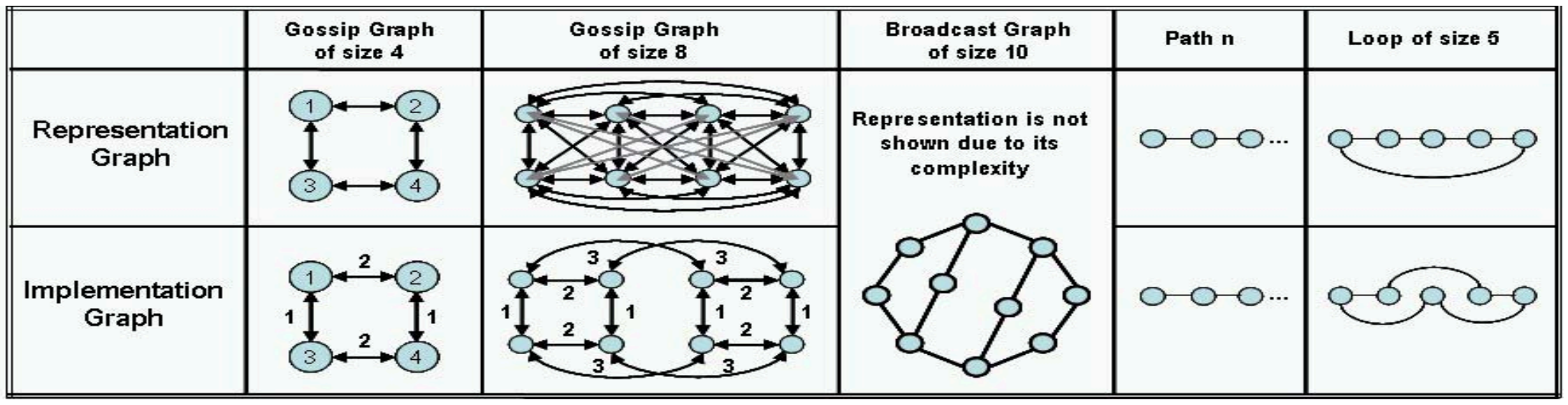

**Figure 1. Sample graphs in the library and their optimal implementations.**

increase and 51% reduction in the energy required to encrypt 128-bit blocks of data in favor of the customized architecture.

The remaining part of the paper is organized as follows: Section 2 reviews related work. The methodology and the implementation of the decomposition algorithm are explained in sections 3 and 4, respectively. Practical considerations and the experimental results appear in Section 5. Finally, Section 6 concludes the paper by summarizing our main contributions.

## 2. Related Work

There has been recent research on finding efficient ways to design NoCs [1-8]. In [8], the authors present a constraint-driven communication architecture synthesis approach based on point-to-point communication specifications. The resulting architecture consists of optimized channels that are obtained by merging or separating the original point-to-point links. Topology selection for application-specific NoCs is discussed in [6]. The authors present a tool that maps the target application to several well-known topologies under various routing scenarios. The resulting designs are evaluated in terms of power, performance and area, and the one giving the best results is selected. Similarly, in [5] the authors specify the application in a simulation environment and generate its dynamic communication graph using simulation traces. Then, the application is mapped to several communication architectures to find the best alternative. Finally, theoretical studies [5-8] report simulation results rather than real implementations, We propose a systematic design methodology for fully customized topology generation, and support our theoretical analysis with direct experiments on a real application.

## 3. Overview of the Proposed Methodology

Generally speaking, it is known how to synthesize the topologies on which the aforementioned communication problems (i.e. broadcasting, gossiping, etc.) can be solved in optimum time. There exists, however, little or no idea about the optimal topology needed for solving a general communication problem. This situation is similar to the logic synthesis problem: The input communication pattern is similar to the uncommitted logic function given to a logic synthesis tool, while the communication primitives appear as standard cells.

To synthesize a customized topology, we decompose the communication requirements of a given application into a set of generic communication primitives, such as gossiping and broadcasting, which may be stored in a communication library. Each primitive in the library has a *representation graph* as shown in Figure 1. This graph structure is the pattern that the decomposition algorithm searches for when processing the input application graph. For example, *gossiping* among 4 nodes implies that each node sends its information to all of the remaining nodes and, at the same time, learns the cumulative message of the network. This is represented by a graph where there is a directed edge from all nodes to all other nodes (see the first graph in Figure 1).

The graphs on which broadcasting (and similarly gossiping) can be completed in minimum time *with minimum number of edges* are called *Minimum Broadcast Graphs* (MBG) and *Minimum Gossip Graphs* (MGG), respectively. These optimal implementations provide the maximum degree of parallelism by enabling the highest number of concurrent communications at minimum cost, as shown in Figure 1 (Note that, in this figure, it is assumed that any processor can participate in at most one communication transaction at any given time instance). These optimal implementations, i.e. MBGs and MGGs, are added as *implementations graphs* to the library. For standard configurations, such graphs are readily available in the literature [10].

After the decomposition step is completed, the communication primitives are replaced by their optimal implementations, and finally glued together to synthesize the customized architecture. As a result, the customized architecture naturally fits the communication requirements of the target applications.

**Design of the Communication Library**

The decomposition algorithm brakes down the input graph into a set of communication primitives stored in a library. Since the final decomposition and the run time of the algorithm itself depend on the primitives in the library, it is desirable to select the best set of graphs to be included in the library. While further research is needed in this area, we construct our current library using the minimum gossip and broadcast graphs that have efficient 2-D implementations and paths and loops of various sizes (Figure 1). This is due to several reasons: First, primitives consisting of a large number of edges will require more wiring resources which is limited by the metal layers allowed for global wires. Second, as the size of the primitives increases, it becomes less likely to detect these primitives in the input graph.

**Energy Characterization of Implementation Graphs**

We assume that, initially, each node in the *implementation graph* holds one bit of information, and stores the path(s) to send this bit to each of the remaining nodes. The energy con-sumed due to the transition of one bit of information from one network node to another node is defined as the *bit energy* ($E_{bit}$) [4] and it is given by

$$E_{bit}^{ij} = n_{hops} \times E_{S\text{-}bit} + (n_{hops} - 1) \times E_{L\text{-}bit} \quad (1)$$

where $n_{hops}$ is the number of hops, and $E_{Sbit}$ and $E_{Lbit}$ are the switch and link energy consumption, respectively. $E_{S\text{-}bit}$ values for different process technologies, voltage levels, operating frequencies are also stored in the library. Modelling $E_{L\text{-}bit}$, on the other hand, requires more attention, since it depends on the link lengths which are not as easy to predict as for the regular grid structures. Hence, we assume that $E_{L\text{-}bit}$ per unit length is stored in the library and the $E_{L\text{-}bit}$ can be obtained from this data given the actual link length and also taking the repeaters into account. Finally, during the execution of the decomposition algorithm, the bit energy definition, the routing information and locations of the cores are used to determine the actual cost of a decomposition, as we describe in Section 4.2.

# 4. Graph Decomposition Algorithm

For simplicity, we assume that the target application is already mapped onto the processing cores and the communication volume between the cores is known. The application is specified by a graph $G(V, E)$, called *Application Characterization Graph (ACG)*, where each vertex represents a core, and the directed edge $e_{ij}$ characterizes the data transfer from vertex *i* to vertex *j*. The communication volume and the required bandwidth from vertex *i* to vertex *j* are denoted by $v(e_{ij})$ and $b(e_{ij})$, respectively. Furthermore, we assume that an initial floorplanning step has been performed and optimized for chip area. Hence, the core coordinates are given as inputs to the algorithm.

## 4.1. Algorithm overview

We propose a depth-first search branch-and-bound algorithm to decompose an arbitrary input graph into a generic set of communication primitives. The algorithm first searches the input (application) graph for a subgraph that is isomorphic to one of the representation graphs in the library. After such a subgraph isomorphism (called a matching) is found, this subgraph is subtracted from the original graph. Subsequently, the same operation is recursively applied to the remaining graph until no matching can be found. For example, in the decomposition shown in Figure 2, the MGG-4 is first identified in the input graph and then subtracted from the input to obtain the remaining graph (see the left most branch in Figure 2). When it is no longer possible to find a subgraph isomorphism in the remaining graph, as is the case in this example, the algorithm stores the remaining graph, traces back to the previous level and continues with the next isomorphism from the library. In this example, a loop of size 4 is detected and the remaining graph looks like the middle branch in Figure 2. Since this graph does not have a subgraph isomorphic to any other graph in the library, the algorithm traces back again to the root of the tree. The final (right most) branch starts with a broadcast graph from one to three nodes. In this case, the remaining graph also has a subgraph isomorphic to one of the representation graphs in the library. Hence, the algorithm goes one level deeper and generates another possible decomposition.

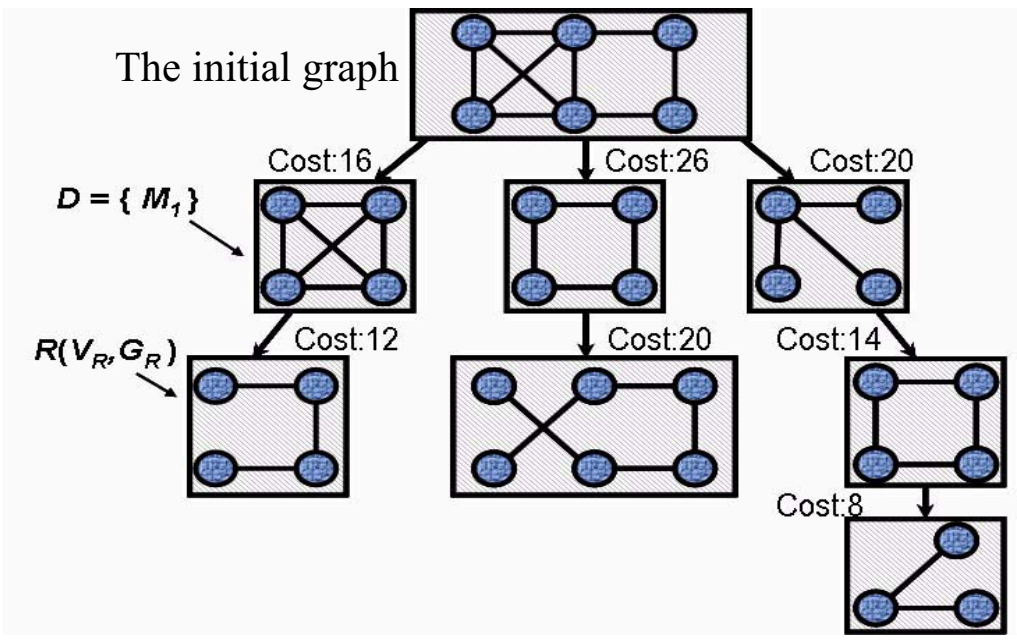

**Figure 2. The illustration of the algorithm.**

In general, there will be more than one possible decompositions. For this reason, we associate a *cost* to each matching and, consequently, obtain a cost for each decomposition as explained in Section 4.3. For instance, in this simple example, the left most branch of cost 16 will be selected as the best decomposition of the initial graph.

## 4.2. Problem Formulation

We consider directed graphs, *G(V,E)*, where *V* is the set of vertices and *E* is the set of edges. The algorithm uses graph addition, subtraction and subgraph isomorphism so, in the following, we provide some basic definitions.

**Definition 1** Given two graphs $G(V_G, E_G)$ and $H(V_H, E_H)$, their *sum* is $A(V, E)$ *such that* $V = V_G \cup V_H$ and $E = E_G \cup E_H$.

**Definition 2** Given a graph $G(V, E)$ and one of its subgraphs $S_G(V_s, E_s)$, their *difference* is called the remaining graph, $R(V_R, E_R)$, such that $V_R = V$ and $E_R = E - E_s$

**Definition 3** A bijective function $f\colon V \to V'$ is *a graph isomorphism* from $G(V, E)$ to $G'(E', V')$ if:

1. For any edge $e = (v_i, v_j) \in E$, there exists an edge $e' = (f(v_i), f(v_j)) \in E'$.
2. For any edge $e' = (v'_i, v'_j) \in E'$, there exists an edge $e = (f^{-1}(v'_i), f^{-1}(v'_j)) \in E$.

An injective function $f\colon V \to V'$ is *a subgraph isomorphism* from $G(V, E)$ to $G'(E', V')$ if there exists a subgraph $S \subseteq G'$ such that *f* is a *graph isomorphism* from *G* to *S*.

**Definition 4** The communication library, $L = \{ P_1, P_2, \ldots, P_n \}$, is the set of representation graphs of the communication primitives. A subgraph isomorphism from the input graph to one of the graphs in the library is called a *matching* and shown as $M\colon V_P \to V_S, P \in L$. Finally, a cost, $C(M)\colon V \times V_P \to R^+$, is assigned to each matching, as explained in Section 4.3.

Given an input graph $G(V, E)$ and a *graph library L,* the decomposition of *G* into *L* is specified by a subset of *L,* called

$D$ and the set of corresponding matchings. The input graph is given by

$$G = \sum_{L_i \in D} M_i(L_i) + R(V_R, E_R) \qquad (2)$$

where $R(V_R, E_R)$ is a remainder graph which does not have a subgraph isomorphism with any communication primitives. The cost of the decomposition is given as

$$C(D) = \sum_{L_i \in D} C(M_i) + C(R) \qquad (3)$$

For example, the cost of the leftmost decomposition, shown in Figure 2, is obtained by adding the costs for the matching $M_1$ and the remainder graph $R$ as 16.

We denote the customized architecture obtained by connecting the implementation graphs of the communication primitives (Figure 1) in $D$, by $I(V_I, E_I)$ and the mapping from the edges of the input graph to the edges in the implementation graphs by $f: E \to E_I$, where $E = \{e_{ij}\}$ is set of edges of $ACG$, and $E_I = \{e^I_{ij}\}$ is set of edges of the implementation graph $I$. Let $\Psi$ be the set of all possible decompositions.

**Problem Statement:** Among all possible decompositions, **find** a decomposition $D^* \subseteq \Psi$, such that the total cost is minimized; that is,

$$C(D^*) = min\{C(D), D \subseteq \Psi\} \qquad (4)$$

subject to the availability of *wiring resources* for the network links and the *bandwidth requirements*; that is,

$$\forall e^I_{ij} \in E_I,\ b(e^I_{ij}) \geq \sum_{e_{ij} \in S} b(e_{ij})$$

where $S = \{e_{ij} | f(e_{ij}) = e^I_{ij}\}$. The former condition is checked by comparing the *bisection bandwidth* of the customized architecture with the maximum bisection bandwidth the particular technology provides for the network links. For example, for matching $M_1$ in Figure 2, edges $e_{13}$ and $e_{14}$ are both mapped to the edge $e^I_{ij}$ in the implementation graph, since if vertex 1 needs to send a message to vertex 4, then it forwards the message to vertex 3 (the first graph in Figure 1). Hence, the bandwidth of $e^I_{ij}$ should be larger than the sum of the bandwidth requirements of $e_{13}$ and $e_{14}$. As a result, the left most branch in Figure 2 will be selected as the best solution provided that the constraints are satisfied.

## 4.3 Cost Assignment

Each path from the root to one of the leaves in the decomposition tree (see Figure 2) constitutes a natural decomposition of the input graph into the primitives. No matter which decomposition is chosen, the maximum number of hops between any two nodes (hence, the average hop number which directly impact the overall performance) in the customized architecture will be bounded by the largest diameter in the communication library. While each decomposition matches the communication requirements of the target application, quantifying the decompositions further is desirable both for selecting the best alternative and for reducing the search space by eliminating certain suboptimal branches. Since our goal is to select the architecture minimizing the total energy consumption, we select the energy consumption as the cost function.

```
I = The input graph;
currentCost = 0;
minCost = inf;
Best Decomposition = NetDecomp(I, minCost, currentcost);

NetDecomp(I, minCost, currentcost)
{
 For all Graphs in the Library G:
    if (a subgraph, S, in I is isomorphic to G) {
       RemainingGraph = I - S;
       currentCost = currentCost + cost of G;
       if (currentCost + minimum remaining cost < minCost)
          ndChild = NetDecomp(I, minCost, currentcost);
       else
          childCost = inf;
       nodeCost = cost of G + min(childCost);
       Check Constraints and Update minCost;
    }
 ndCost = Cost of the Remaining Graph; // None of the graphs
 in the library match to the input.
```

**Figure 3. The pseudo-code of the graph decomposition algorithm.**

The energy consumed due to the transportation of one bit of information from network node $i$ to node $j$ is given in Equation 1. Since the vertices of the $ACG$ specify the communication volume between each pair of vertices, we can compute the *total energy consumption of a matching M* as

$$C(M) = \sum_{e_{ij} \in M_{imp}} (E^{ij}_{bit}(l_{ij}) \times v(e_{ij})) \qquad (5)$$

where $M_{imp}$ is the set of edges in the implementation graph of the library component. Note that $E_{bit}$ is a function of the link length, $l_{ij}$. Since we assume that the positions of the cores are determined by an initial floorplaning stage, the distances between all vertex pairs are known a priori. Hence, accurate $E_{bit}$ values can be imported from the library.

## 4.4 Details of the Decomposition Algorithm

The goal of the decomposition algorithm is to cover the input $ACG$ with the set of library graphs resulting in the minimum total cost. We solve this minimization problem using a branch-and-bound algorithm as shown in Figure 3.

Initially, the cost of the covering is set to zero and the minimum cost achieved so far is set to infinity. The algorithm proceeds as explained in Section 4.2. The cost of each matching is computed using Equation 5, and when a complete decomposition is found (i.e. the algorithm reaches a leaf node), the cost of the decomposition is calculated with Equation 3. If this cost is smaller than the minimum cost obtained so far, the minimum cost is updated. In order to bound the search, we check the current cost of a decomposition and the minimum possible cost decomposing the remaining graph. If their sum is larger than the current minimum cost, the algorithm marks the cost of this branch as infinity and traces back to the previous level. Finally, the legal decomposition with minimum cost is selected as being the best decomposition.

### 4.5. The Routing Strategy

While the designer has complete freedom to select any deadlock-free routing strategy, we generate a routing table as a by-product of the topology synthesis algorithm using the following observation. We know the optimal strategies that allow broadcasting (also gossiping) on MBGs (MGGs) [10,11]. The numbers on the implementation graphs, in Figure 1, show how gossiping can be completed in minimum number of rounds. For example, for MGG-4, the nodes (1,3) and (2,4) exchange their information during the first round. After that, during the second round, nodes (1,2) and (3,4) exchange the information they learned so far. Hence, after two rounds every node knows the initial information of all other nodes. Knowing this optimal schedule, we can generate a routing table such that each vertex knows precisely how to send a message to the vertices it is not directly connected to in the implementation graph. In this example, if vertex 1 needs to send a message to vertex 4, then it will forward its message to vertex 3 first, since there exists an optimal schedule which delivers the information to vertex 4 using this route.

Each node in the *ACG* holds a table containing the nodes to which it can send a packet. This table is updated during the decomposition process such that the nodes keep track to which neighbor they should forward a given packet. The cycles that can cause deadlock can be detected and avoided by the algorithm, while it is also possible to eliminate such cycles by introducing virtual channels.

## 5. Experimental Results

### 5.1. Experiments with Random Graphs

A set of benchmarks generated using TGFF [17] and Pajek [14] are used to evaluate the run time of the algorithm and illustrate the decomposition approach.

The largest run time obtained for TGFF is 0.3 seconds corresponding to an automotive industry benchmark consisting of 18 nodes, as shown in Figure 4a. Figure 4b shows the average run time obtained using more than 60 larger graphs generated by Pajek. While the average run times are longer for this case, the algorithm can still decompose a graph with 40 nodes in less than 3 minutes.

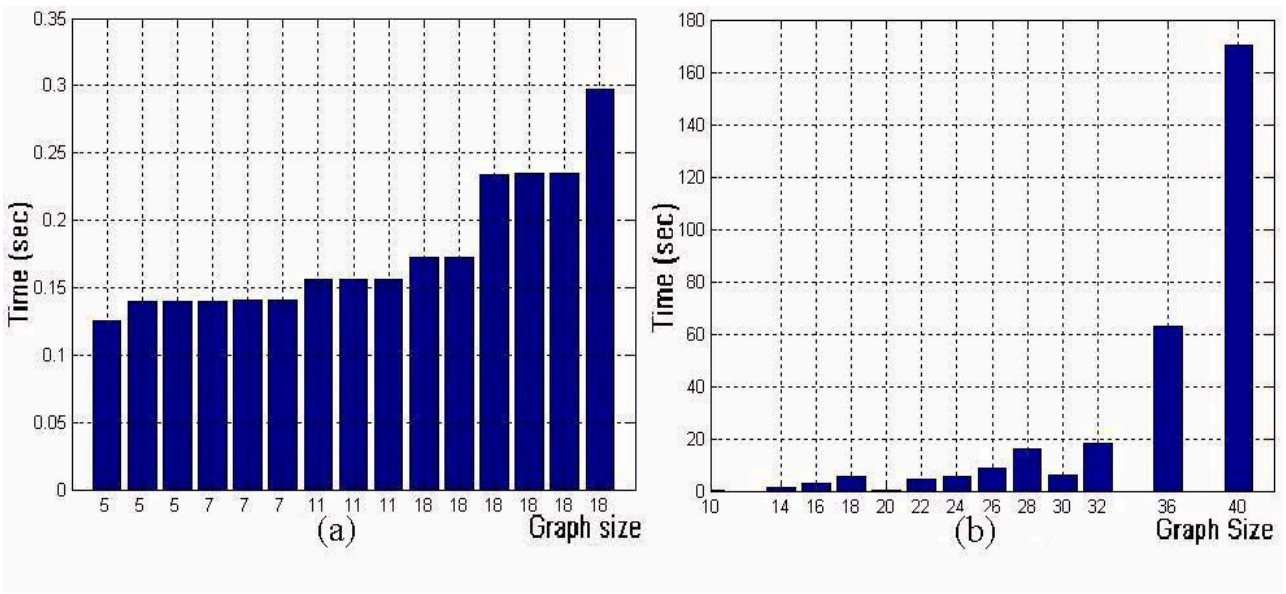

**Figure 4. Run time of the algorithm for random graphs generated by TGFF (a) and Pajek (b)**

The current version of the proposed approach is implemented in Matlab. The branch-and-bound algorithm calls a subgraph isomorphism function based on VF2 [13] written in C++. The run-time of the algorithm can be improved in a number of ways. While the VF2 isomorphism algorithm works simultaneously on two graphs, approaches that consider a collection of model graphs (like our library) and generate a decision tree for faster identification of subgraph isomorphisms have been developed [15]. Furthermore, the requirement for perfect matching can be relaxed and the graphs that are sufficiently close to each other can be detected [16]. Finally, the run time can increase drastically, if the input graph does not have a subgraph isomorphic to the graphs in the library, because the algorithm tries all different permutations before quitting. Hence, the search for the isomorphism can be terminated after a time-out period rather than trying all permutations.

**Example**: In order to illustrate the architecture synthesis process more clearly, we show one randomly generated *ACG* (using Pajek) and its customized implementation as in Figure 5. While the communication patterns in the input graph are not easily detectable by eye inspection, the algorithm decomposes it into the primitives in less than 0.1 seconds:

```
1: MGG4,   Mapping: (1 1), (2 2), (3 5), (4 6)
  3: G123,     Mapping: (1 3), (2 2), (3 5), (4 6)
    3: G123,     Mapping: (1 7), (2 3), (3 5), (4 6)
      2: G124,     Mapping: (1 8), (2 1), (3 3), (4 6), (5 7)
        3: G123,     Mapping: (1 4), (2 5), (3 6), (4 7)
```

The output starts with the ID of the communication primitive in the library and its label. For example, in the sample output shown above, the first match is a *gossip graph* of size 4, whose ID in the library is 1. The following three matches are *broadcast graphs* from one node to three nodes and the final match is another *broadcast graph* from one node to 4 four nodes. In this particular example, there is no remaining graph after these matches are found. The algorithm also outputs the mapping from the graph in the library to the isomorphic subgraph. For example, the mapping in the first line shows that the vertex 1 of MGG4 is mapped to *v1* in the input graph. Similarly, vertices 2,3 and 4 are mapped to *v2*, *v5* and *v6*, respectively. If we investigate these vertices in the input graph (Figure 5a), we can indeed observe the gossip graph shown in Figure 1.

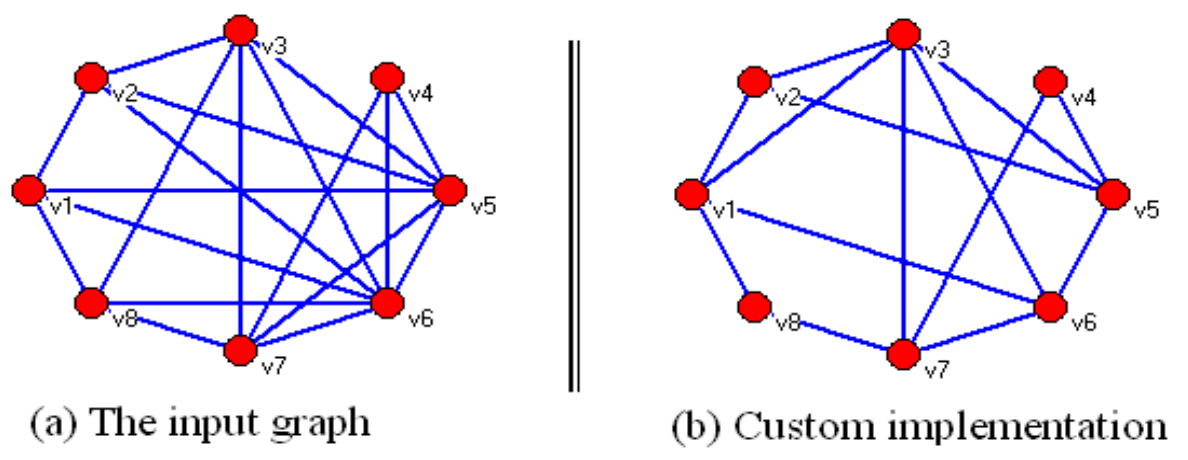

**Figure 5. The illustration of the customized synthesis for a random benchmark.**

### 5.2. Distributed Implementation of AES

We distributed the AES operations to a network of 16 identical nodes each processing one byte of the input block and obtained the application characterization graph shown in Figure 6a. Then, we generated a customized communication architecture using the proposed algorithm. The algorithm found the following decomposition in 0.58 seconds:

```
COST: 28
  1: MGG4,     Mapping: (1 1), (2 5), (3 9), (4 13)
   1: MGG4,     Mapping: (1 2), (2 6), (3 10), (4 14)
     1: MGG4,     Mapping: (1 3), (2 7), (3 11), (4 15)
       1: MGG4,     Mapping: (1 4), (2 8), (3 12), (4 16)
         2: L4,        Mapping: (1 5), (2 6), (3 7), (4 8)
           2: L4,        Mapping: (1 13), (2 14), (3 15), (4 16)
               0: Remaining Graph:
```

This decomposition shows that the algorithm successfully captures the all-to-all communication patterns within the columns of *ACG*. The first line of the output shows that the vertices 1, 2, 3, 4 of the library graph is mapped to the vertices 1, 5, 9, 13 of the input graph, which is the first column. Similarly, the other columns are also mapped to a gossip graph of size 4. The output also shows that the second and fourth rows are mapped to loops of size 4. Finally, the remaining subgraph after these matches (the third row) cannot be matched to any graph in the library. Hence, it is reported as the remaining graph.

The resulting architecture (Fig. 6b) and the standard mesh architecture have prototyped using a Virtex 2 based development board equipped with a XC2V4000 device. Both designs utilize roughly 32% of the device resources.

### Prototype Performance and Energy Comparison

The chip *throughput* and average *latency* experienced by the packets in the network are utilized to measure and contrast the performance of the proposed architecture and the standard mesh architecture. The throughput is expressed as

$$\rho = 128\frac{bits}{block} f_{clk} / \left(\Delta\frac{clk}{block}\right)$$

where Δ cycles/block is the time needed to encrypt one block (128 bits) of input data. For the mesh configuration, we measured directly on the prototype $\Delta = 271$ *cycles/block* which gives $\rho = 47.2$ Mbps at a clock frequency of 100Mhz. On the other hand, for the customized topology, we measured $\Delta = 199$ *cycles/block*s, resulting in a increased throughput of 64.3 Mbps. Similarly, there is a 17% reduction in latency. The mesh network causes 11.5 cycle average latency, while for the customized architecture the average latency is only 9.6 cycles.

We also measured the *power* consumption of the two designs with the *Xpower* utility of Xilinx, after placement and routing, using the actual simulation traces. We measured 33% reduction in the average power consumption compared to a standard mesh architecture. The energy consumed per 128-bit input block is the product of the time it takes to encrypt the block and the average power consumption during this period, i.e., $E = (\Delta/f)P_{ave}$. As such, the mesh architecture requires $5.1\mu J$, while the customized architecture requires only $2.5\mu J$ to encrypt one block; this results in roughly 51% energy savings.

The AES algorithm enables us to demonstrate our approach and develop NoC prototypes based on customized and standard mesh implementations. However, due to its modest processing and communication requirements, AES is far from demonstrating the benefits of a networked implementation.

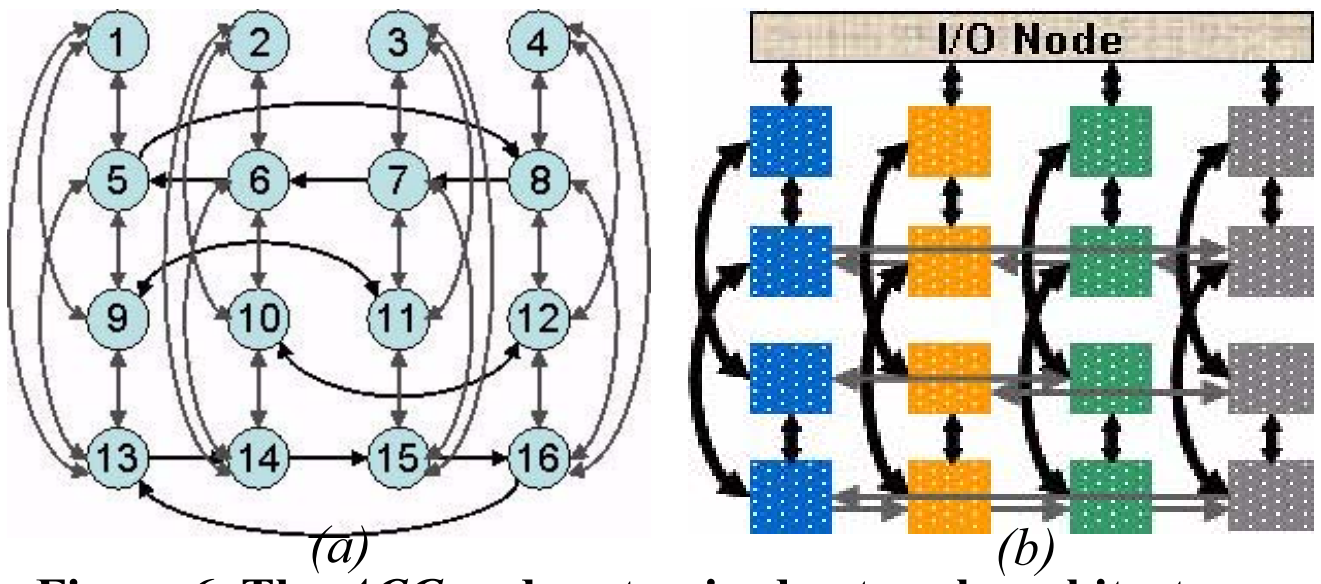

**Figure 6. The *ACG* and customized network architecture.**

## 6. Conclusion and Future Work

In this paper, we presented a methodology for customized communication architecture synthesis. The communication requirements of the algorithm are decomposed into a set of frequently encountered communication primitives using an efficient branch-and-bound algorithm. The algorithm searches efficiently the entire design space for the architecture that minimizes the total energy consumption of the system while satisfying the other design constraints. The effectiveness of our methodology is illustrated using the AES algorithm and some random graphs.

There are several future research directions. For instance, it is possible to relax the initial floorplan information and solve the optimization problem for the general case. Also the possibility of using adaptive or stochastic routing strategies should be investigated.

## 7. References

[1] W. Dally and B. Towles, *"Route packets, not wires: On-chip interconnection networks,"* In Proc. 38th DAC, June 2001.

[2] A. Hemani, *et al., "Network on Chip: An Architecture for Billion Transistor Era,"* In Proc. IEEE NorChip Conf., Nov. 2000.

[3] P. Guerrier, A. Greiner, *"A Generic Architecture for On-Chip Packet Switched Interconnections,"* In Proc. DATE, March 2000.

[4] J. Hu, R. Marculescu, *"Exploiting the Routing Flexibility for Energy/Performance Aware Mapping of Regular NoC Architectures,"* In Proc. DATE, March 2003.

[5] M. Kreutz, *et. al.* "*Communication Architectures for System-On-Chip,"* In 14th Symp. on Integrated Circuits and Systems Design, Sep. 2001.

[6] S. Murali, G. De Micheli, *"SUNMAP: A Tool for Automatic Topology Selection and Generation for NoCs,"* In Proc. 41st DAC, June 2004.

[7] A. Jalabert, *et. al.* "*xpipesCompiler: A Tool for instantiating application specific Networks on Chip*," In Proc. DATE, March 2004.

[8] A. Pinto, L. P. Carloni, A. L. Sangiovanni-Vincentelli, *"Efficient Synthesis of Networks On Chip,"* In Proc. ICCD, Oct. 2003.

[9] http://csrc.nist.gov/CryptoToolkit/aes/rijndael/

[10] S. M. Hedetniemi, S. T. Hedetniemi and A.L. Liestman, "A survey of gossiping and broadcasting in communication networks," Networks 18(4), 319-359 (1988).

[11] J. Hromkovic, *et. al.*, *"Dissemination Of Information In Interconnection Networks (Broadcasting & Gossiping),"* Combinatorial Network Theory, D.-Z. Du and D.F. Hsu (Eds.), Kluwer Academic Publishers, Netherlands, 1996.

[12] L. P. Cordella, *et. al.*, *"A (Sub)Graph Isomorphism Algorithm for Matching Large Graphs,"* IEEE Trans. on Pattern Analysis and Machine Intelligence, Oct. 2004.

[13] ARG Database CD, http://amalfi.dis.unina.it/.

[14] http://vlado.fmf.uni-lj.si/pub/networks/pajek/

[15] B. T. Messmer, H. Bunke, *"A Decision Tree Approach to Graph and Subgraph Isomorphism Detection,"* Pattern Recognition, Dec. 1999.

[16] P. Foggia, *et al.*, *"A Performance Comparison of Five Algorithms for Graph Isomorphism,"* In Proc. 3rd IAPR TC-15 Workshop on Graph-based Representations in Pattern Recognition, May, 2001.

[17] R. P. Dick, *et al.*, *"TGFF: Task Graphs For Free,"* Proc. Intl. Workshop on Hardware/Software Codesign, March 1998.